\documentclass[nofootinbib,aps,pre,showpacs,twocolumn]{revtex4}

\usepackage{amsmath}
\usepackage{amssymb}
\usepackage{graphicx}

\newcommand{\EXP}[1]{ \mbox{\large e}^{#1}}   
\newcommand{\imat}{{\mathrm{i}}}     
\newcommand{\hbare}{\hbar_{\mathrm{eff}}}

\begin{document}

\author{Amaury Mouchet}
\email{mouchet@celfi.phys.univ-tours.fr}
\affiliation{Laboratoire de Math\'ematiques et de Physique Th\'eorique, 
             Universit\'e Fran\c{c}ois Rabelais,
             Ave\-nue Monge, Parc de Grandmont 37200
             Tours, France.}
\homepage{http://www.phys.univ-tours.fr/~mouchet}
\author{Dominique Delande}
\email{delande@spectro.jussieu.fr}
\affiliation{Laboratoire Kastler-Brossel, 
             Universit\'e Pierre et Marie Curie, 
             4, place Jussieu, F-75005
             Paris,  France. }

\title{Signatures of chaotic tunnelling}

\date{\today}

\begin{abstract}
Recent experiments with cold atoms provide a significant
step toward a better understanding of tunnelling when irregular
dynamics is present at the classical level. In this paper, we lay out
numerical studies which shed light on the previous experiments, help
to clarify the underlying physics and have the ambition to be
guidelines for future experiments.
\end{abstract}


\maketitle

\bigskip
  \section{Introduction} 

When studying tunnelling in non-separable systems with more
than one degree of freedom, one immediately encounters difficulties
which generically can be traced back to the absence of enough
constants of motion.  Even in the very peculiar case of integrable
systems, where continuous symmetries provide as many constants of
motion as degrees of freedom, as soon as separability is lost,
the
analysis of tunnelling is not a simple generalization of what occurs
in one-dimensional (1D) autonomous systems. 
The later case is detailed in textbooks on quantum physics (see for
instance~\cite{Messiah65a}) and it has even been possible to give a
comprehensive analytical treatment in term of complex solutions of
the Hamilton equations~\cite{Balian/Bloch74a}. However, it is not
until the mid-eighties that a satisfactory quantitative approach has
been proposed~\cite{Wilkinson86b,Wilkinson/Hannay87a} for tunnelling
in non-separable integrable systems involving a larger number of
dimensions.  Moreover, integrability is a property of higher
dimensional systems which is not generic.  The coupling between
several internal degrees of freedom as well as the coupling to an
external source usually
destroys some global constants of motion. With such a lack of
constraints on the dynamics, the classical motion in phase space 
may become chaotic: it may explore volumes with higher
dimensionality and therefore becomes exponentially sensitive on
initial conditions.  It is not surprising that these deep qualitative
differences between an integrable regime and a chaotic one 
appear at the quantum level too. Some of the properties of a
\emph{quantum} system do change when constants of motion are
broken and, indeed, it is the very object of quantum chaos to 
study the signatures of classical chaos at the quantum level (see
for instance~\cite{Giannoni+91a} to see how rich, vivid and
successful this domain is).

We define tunnelling as a quantum process
which is classically forbidden to \emph{real} solutions of classical
equations of motion. In this paper, we 
consider Hamiltonian  systems only and study how the 
non-dissipative breakdown of continuous symmetries
affects tunnelling. We do not consider how tunnelling is modified by 
dissipation and decoherence of the quantum wave. 
Of course,
this requires a great care in real experiments where making
dissipation negligible is always a hard task. This is
one of the main reasons why very few real experiments have been done on
these questions, though they would definitely help to understand 
tunnelling in the presence of chaos (as far as we know, the only
experiments explicitly made on chaotic tunnelling in the XX$^{\rm{th}}$ 
century are those presented in~\cite{Dembowski+00a} 
with electromagnetic microwaves
instead of quantum waves).  

During the last fifteen years however,
theoretical and numerical investigations on autonomous 2D and
time-dependent 1D Hamiltonians systems  highlighted some
mechanisms~\cite{Wilkinson86b,Wilkinson/Hannay87a,Lin/Ballentine90a,%
Bohigas+93a,Bohigas+93b,Tomsovic/Ullmo94a,Creagh/Whelan96a,Brodier+02a}
and much insight has been gained on the influence of 
non-separable dynamics. Experimental evidence of such mechanisms would
be of great interest especially in the light of the subtle interplay
between interferences and disorder.  These phenomena lie in
the general context of wave transport in complex media where
the role of disorder is played by the (deterministic) chaotic dynamics
instead of having a statistical random origin. Of course, other
important motivations can be found in the numerous domains where
tunnelling plays a crucial role as a fundamental quantum process like
ionization~\cite{Zakrzewski+98a}, absorption, nuclear
radioactivity, molecular collisions, mesoscopic physics etc. 
More speculatively,
studies on tunnelling in high dimensional
Hamiltonian systems should provide us with a natural extension of the
instanton techniques which  deal with
quantum field theories reducible to effective 1D
autonomous Lagrangian systems.
 
In 2001, it has been shown both theoretically~\cite{Mouchet+01a} and
experimentally~\cite{Hensinger+01a,Steck+01a,Hensinger02a,Hensinger+02a} that atom cooling
techniques~\cite{Arimondo+92a} (and possibly molecular physics as
well, where formally similar systems have been extensively
studied~\cite{Lawton/Child81a,Sibert+82a,Kellman/Lynch87a,Rose/Kellman96a})
provide systems which fulfill all the severe requirements to study
tunnelling in the presence of classical Hamiltonian chaos: accurate
manipulation of internal and external degrees of freedom, precise
control of  dissipation and decoherence and on the preparation/detection set
up. For a brief account intended for a large audience
see~\cite{Heller01a,Mouchet/Ullmo01a,GossLevi01a}).  The present article 
has the ambition to mark out the
future challenging experiments, that remain to be done for reasons which
will hopefully appear clear in the following.

This paper is organised as follows. In
section~\ref{sec:chaotictunnelling} we  give a general and
informal overview.  In
section~\ref{sec:effhamsyst}, we quickly recall the main theoretical
apparatus that is needed in the following.  We implicitly refer
to~\cite{Mouchet+01a} for details and demonstrations.  In
section~\ref{sec:experiments}, we  comment the results
of~\cite{Hensinger+01a} and~\cite{Steck+01a}. In
section~\ref{sec:numexp}, we show in this context, with the help
of numerical experiments, the very precise form taken by the phenomenon
known as chaos assisted tunnelling. We explain why it has not
been observed yet with real atoms and propose how to actually bring it
to the fore.  Before the concluding remarks in
section~\ref{sec:conclusion}, we give in
section~\ref{sec:subtleties} some more numerical results which 
illustrate how subtle the signatures of chaotic tunnelling can be.

\section{Chaotic tunnelling}
\label{sec:chaotictunnelling}
\begin{figure}[!ht]
\center  
\includegraphics[width=8.5cm,keepaspectratio]{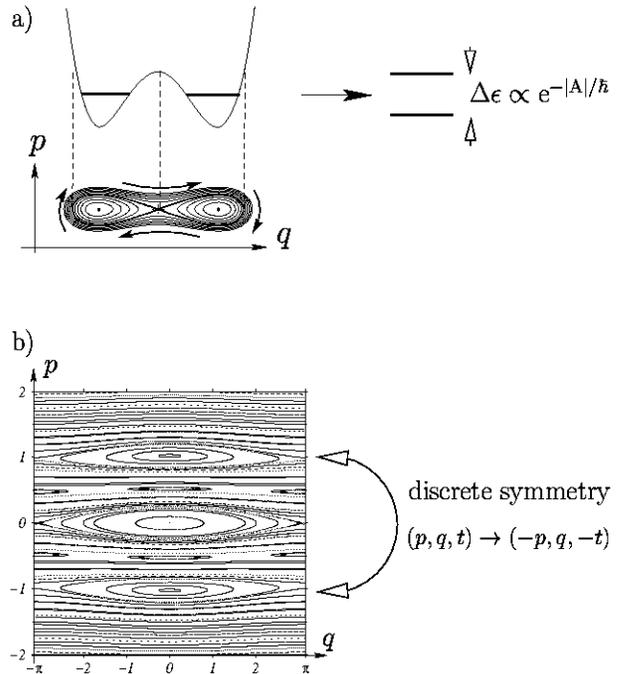}
\caption{\label{fig:2well_generalization1}
A generalization of the paradigmatic double well potential (a) is to consider tunnelling between
stable islands that are related by any discrete symmetry in phase space. Case (b) corresponds
to Hamiltonian~\eqref{eq:ham}
with $\theta=1$ and $\gamma=0.018$. 
Here, the time reversal symmetry plays the role of parity in case (a). 
}
\end{figure}

The simplest situation to illustrate tunnelling is probably the case
of a particle placed in a 1D time-independent symmetric
double-well potential (see Fig.~\ref{fig:2well_generalization1}~(a)).
Starting in one
well with an energy that is below the maximum of the
potential, a quantum particle can jump into the other well with a non zero probability
though it is a forbidden classical process. In addition to the
classical time scale~$\tau$ given by the oscillating period
\emph{inside} one well, we therefore have a longer 
time scale, the tunnelling period~$T\gg\tau$ of
the oscillations \emph{between} the wells. In the eigenenergy spectrum,
tunnelling appears as a quasi-degeneracy of the odd and even-symmetry
states whose energies are both of the order of~${\hbar}/{\tau}$ but
differ by an exponential energy splitting
\begin{equation}\label{eq:wilkinson}
	\Delta \epsilon= \frac{2\pi\hbar}{T}\sim\EXP{-A/\hbar}
\end{equation}  
where~$A$ is a $\hbar$-independent typical action
and can be interpreted in terms of a unique
complex classical trajectory under the barrier~\cite{Messiah65a,Heading62a}.

In the following we generalize this elementary situation in two
ways.  First, unlike the parity in the previous example, we can deal
with a symmetry which is not necessarily either a spatial one or a
two-fold one.  In other words, we can have any discrete symmetry group acting
on the whole phase space as well as any $N$-fold symmetry which
lead to bunches of $N$-uplets in the energy spectrum (or bands if
$N\gg1$).  In the following we keep $N=2$ since we 
have a two-fold symmetry~$\mathcal{T}$ actually playing the role of
parity (see Fig.~\ref{fig:2well_generalization1}~(b)) 
and  being somehow decoupled from
the other discrete symmetries.  The classical structure in phase space
is globally invariant under~${\mathcal{T}}$ 
and the quantum eigenstates can be classified according to
their symmetric or antisymmetric character under the unitary
transformation which represents~$\mathcal{T}$ in the Hilbert space of
states.  
Because $\mathcal{T}$ acts in phase space, it is usually more complicated
than a pure spatial
transformation. Thus, the two regions of phase
space connected by {\em quantum} tunnelling, but {\em classically} not
connected, are in general not separated by a simple potential barrier,
but by a more complicated dynamical barrier.  
In such a case, tunnelling
is emphasized to be called ``dynamical tunnelling'' as suggested by
Davis and Heller in~\cite{Davis/Heller81a}. It often happens that
the classically unconnected region are associated to the same
region of configuration space, with different momenta. A simple study
of the density probability in configuration space is then unsufficient
to characterize dynamical tunnelling; an analysis of the density probablility in 
momentum  space is required.

\begin{figure}[!ht]
\center  
\includegraphics[width=8.5cm,keepaspectratio]{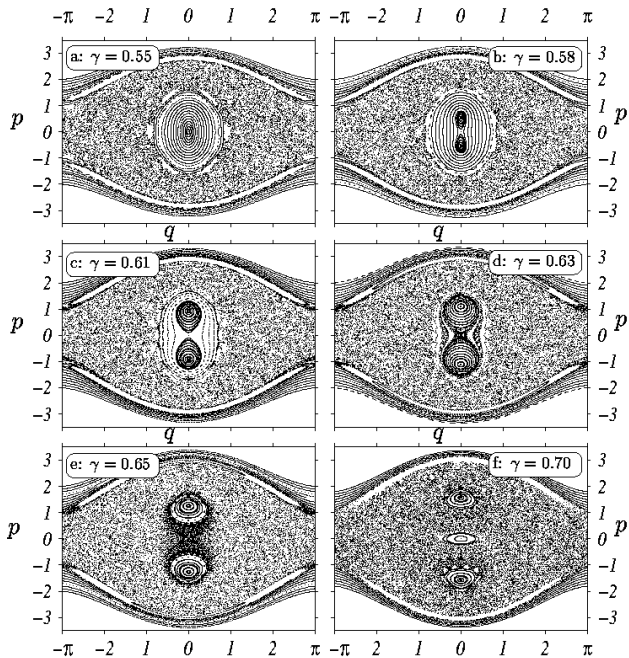}
\caption{\label{fig:bifurc2yeuxclassique} Poincar\'e surfaces of section 
correspond to Hamiltonian~\protect{\eqref{eq:ham}}
with $\theta=1.724137$. 
Two stable islands in the vicinity of the origin
are created by a pitchfork bifurcation at~$\gamma\simeq0.56$. Above
this value, chaotic motion progressively invades
phase space in between the two stable islands. 
At~$\gamma>\gamma_c\simeq0.625$, the latter
are no longer connected by regular trajectories. 
The experimental configuration used in the \textsc{nist} experiments
corresponds
to~$\gamma=0.96$ just before the islands disappear 
in a bifurcation cascade at~$\gamma\simeq0.97$.
}
\end{figure}

The second kind of generalization leads to much more puzzling
questions. When dealing with systems with several degrees of freedom or,
equivalently, if an external time dependence exists, 
classical trajectories generically lose
their regular behavior, cannot analytically be computed and are
organized in a fractal hierarchy that is described by the 
KAM perturbative scenario.  Recently,
important progress has been achieved in the 
understanding of the continuation of these
intricate structures in complex phase space and
their role at the quantum level, see~\cite{Kus+93a,Leboeuf/Mouchet99a}
and especially~\cite{Shudo/Ikeda98a}.  Anyway, we are therefore
led to the following typical quantum chaos question: if
one is able to create  two symmetric
stable islands separated in classical phase space 
by a chaotic sea whose volume is under
control (see Fig.~\ref{fig:bifurc2yeuxclassique}), what is the effect 
of this sea on
the (dynamical) tunnelling between the islands?

The
``dual'' situation where chaos is created \emph{inside} the wells
while the dynamical barrier is kept regular has been introduced and
studied theoretically and numerically in~\cite{Creagh/Whelan96a}.  For
a better understanding of what occurs in the energy spectrum when
regular wells are separated by a chaotic sea, it has been proposed
in~\cite{Tomsovic98b} to slightly break the tunnelling
symmetry. Nevertheless, in the following of the present paper, it
must be kept in mind that a discrete symmetry will be always
maintained exactly.  At last, a third kind of generalization where the
Hamiltonian character is destroyed by introducing dissipation and/or
coupling to a thermal bath, is beyond the scope of this work~\cite{Tomsovic98a}.

\section{Effective Hamiltonian system}
\label{sec:effhamsyst}

Following~\cite{Mouchet+01a,Hensinger+01a,Steck+01a} (see
also~\cite{Averbukh+95a} in a different context) we deal with an
effective 1D time-dependent system whose Hamiltonian is
\begin{equation}\label{eq:ham}
H(p,q;t)
        =\frac{p^2}{2}
        -\gamma \ (\theta + \cos t )\ \cos q
\end{equation} 
 in dimensionless units. $\gamma$ and $\theta$ are two classical real
parameters that can be modified in real experiments.  In addition,
there is also one parameter, namely $\hbare$, which fixes the quantum
scale and is defined by the usual relation between canonical
operators:~$[q,p]=\imat\hbare$. It turns out that $\hbare$ is not
constant any longer 
(see section~\ref{sec:experiments} below).
It can be also experimentally varied via the
rescaling factor that is needed in the canonical commutation relation
in order to work in dimensionless units used to
write~\eqref{eq:ham}.  

The time dependence breaks the
conservation of energy and therefore may generate chaos.  
In order to deal with such a Hamiltonian, it is crucial
to remark that it has both a spatial and a temporal periodicity.
The latter implies that the Floquet theorem can be used, 
which states that the Hilbert space is spanned by an orthonormal
eigenbasis of the evolution operator over one period. The
corresponding eigenvalues of this unitary operator are distributed on
the unit circle and therefore are labelled by their phase,
which is conveniently written as $\exp{(\imat2\pi \epsilon/\hbare)},$
where $2\pi$ stands for the period of the modulation and
$\epsilon$ can be interpreted as a quasi-energy, a generalization of
the notion of energy level for a time-periodic system.

The spatial periodicity of the Hamiltonian is also extremely
important, as it makes it possible to split the Hilbert space 
into independant components, each component being characterized
by the so-called Bloch vector $k$ in the $]-0.5,0.5]$ range: 
under translation of $2\pi$ along
$q,$ the quasi-energy eigenstates are just
multiplied by the phase factor $\exp{(\imat2\pi k)}.$
One is thus reduced to solve the Floquet-Schr\"odinger equation
in an elementary spatial cell with boundary conditions depending
on $k.$ One thus generates -- for a fixed value of $k$ -- a discrete
quasi-energy spectrum $\epsilon_i(k).$ When the full range of $k$ values
is considered, one obtains the familiar 
(quasi-)energy bands~\cite{Ashcroft/Mermin76a}.

There is an additional discrete symmetry which can be used.
The Hamiltonian~\eqref{eq:ham} is invariant under the time-reversal
symmetry $(q,p,t)\to(q,-p,-t).$ In the classical
surfaces of section, this implies a symmetry with respect to the
$q$ axis. In situations like the one in fig.~\ref{fig:2well_generalization1}b,
this implies the existence of pairs of symmetric classically
unconnected tori, i.e. a situation where tunnelling
could be observed. 
In the quantum world, the situation is slightly more
complicated, because this symmetry connects the $k$ subspace
to the $-k$ subspace. In the particular case $k=0$ ($k=0.5$ could
also be used), this implies that the Floquet eigenstates can be
split in two subclasses of states which are either even or odd
under the symmetry operation. The 
splitting between a doublet
of even and odd states, 
$\Delta \epsilon_n=|\epsilon_n^{+}(0)- \epsilon_n^{-}(0)|$
will be a measure of tunnelling.

We will extensively use the Husimi
representation of a quantum 
state~\cite{Husimi40a}.
  Such a
representation associates to each quantum state $|{\psi}\rangle$ a
phase space function~$\psi^{\mathrm{\scriptscriptstyle H}}(p,q)$
(where $p$ and $q$ are real numbers) defined by
\begin{equation}
\psi^{\mathrm{\scriptscriptstyle H}}(p,q) =|\langle{z}|{\psi}\rangle|^2        
\end{equation}
where~$|{z}\rangle$ is the  coherent state corresponding to
the complex number~$z=(q+\imat p)/\sqrt{2\hbare}$. 
Because~$|{z}\rangle$ is a
minimal Gaussian wave packet with average momentum $p$ and average
position $q$, 
the Husimi function $\psi^{\mathrm{\scriptscriptstyle H}}(p,q)$ 
contains some information about the degree of localization
of~$|{\psi}\rangle$ in phase space, and makes it possible to
associate quantum states with classical phase space structures.
 
\section{Experiments with cold atoms}
\label{sec:experiments}

Under some severe conditions which constrain the experiments, 
Hamiltonian~\eqref{eq:ham} can be obtained as an effective dimensionless 
Hamiltonian for cold neutral independent atoms of mass~$M$ interacting with
two counterpropagating laser beams~\cite{Mouchet+01a,Hensinger+01a,Steck+01a}. 
These two beams  have two
slightly different frequencies at~$\omega_L+\delta\omega/2$ and~$\omega_L-\delta\omega/2$.
The longitudinal coordinate~$x$ and the dimensionless~$q$ are related by
$q=2k_Lx$ where $k_L=\omega_L/c$. The rescaling of the momentum is given
 by~$p=(2k_L/M\delta\omega)p_x$.
 $\gamma$ and~$\theta$ are fixed by the  intensity of the lasers and the detuning of
 the laser frequencies with respect to the atomic resonance.
The dimensionless time~$t$ is taken in $\delta\omega^{-1}$ units and the expression of the
effective 
Planck constant is~$\hbare=8\omega_R/\delta\omega$
where $\omega_R=\hbar k_L^2/2 M$.
 Since \textit{in
fine} we want to measure exponentially small tunnelling
splittings~$\Delta\epsilon$, it requires to maintain these conditions
for a time at least larger than $\hbare/\Delta\epsilon$.
Moreover, a very accurate control of the preparation of the initial
state and of the analysis of the final state is compulsory.

As shown above, because of the temporal and spatial periodicity 
of the Hamiltonian, observing the standard signature of tunnelling,
that is an oscillation of a quantum state between two classically
unconnected regions of phase space requires that a single doublet
of Floquet-Bloch eigenstates is initially populated, with well
defined values of the  parameters $(\gamma,\theta,\hbare)$
but also with  a well defined value of the Bloch angle $k$. 
If more than a single doublet is populated, additional
frequencies (related to energy
differences between the various populated Floquet states)
will appear in the temporal evolution. If any parameter is not fixed,
the experimental signal will be the superposition of tunnelling
oscillations (with different frequencies) for various sets of parameters.
This will at best -- if the dispersion of the parameter values is
reasonnably small -- blur the oscillations at long time and at
worst will completely destroy the signature of dynamical tunnelling.
It is experimentally rather simple to keep an accurate time-periodicity
of the driving signal, i.e. to fix $\hbare.$ Similarly, the balance
between the constant and the oscillatory term, hence the
parameter $\theta$ is easily controlled. The $\gamma$ parameter is proportional
to the laser intensity and may thus slightly vary across the atomic cloud
(because of the transverse structure of the laser beams).
The most difficult part is to control that a single Bloch angle $k$
is excited. Indeed, this requires a phase coherence of the initial wavefunction
over a large number of laser wavelengths, which is extremely
difficult to achieve experimentally~\cite{Greiner+02a}, as will be shown in the following.
In any case, the inhomogeneous broadening of the experimental signal
because of the dispersion in $k$ will be responsible for a decay
of the tunnelling oscillations.

\subsection{NIST experiments~\protect\cite{Hensinger+01a}}
\label{subsec:Hensinger}

In the \textsc{nist} experiments, the two stable symmetric islands are
chosen quite close in phase space in order to deal with not too small
splittings. 
Another crucial point of this experiment
is that the classical motion of the
islands over one period, unlike those in~\cite{Mouchet+01a} and~\cite{Steck+01a},
always remains trapped in one spatial elementary cell of length
$2\pi$. The quantum states localized in
these islands are consequently only weakly sensitive 
on the boundary conditions
which are governed by the Bloch angle. In other words,
the tunnelling period will be only 
weakly dependent on the Bloch angle $k$, which 
implies that the unavoidable broadening over $k$ will not 
spoil too much the signature of tunnelling. This is
a major improvement over the
tunnelling described in~\cite{Mouchet+01a} and~\cite{Steck+01a},
where a very narrow band of Bloch angle is required
to observe clear tunnelling oscillations. Moreover, the atoms involved in the 
tunnelling process stay longer in the
region where the laser intensities are 
uniform.

Indeed, as proposed in~\cite{Hensinger+01b}, \cite[chap. 4 and 5]{Hensinger02a} 
the two stable symmetric islands are created from a
pitchfork bifurcation of the fixed point at $(p,q)=(0,0)$. To
visualize it (see Fig.~\ref{fig:bifurc2yeuxclassique} (a) and (b)), we
extract a one-parameter sequence by varying $\gamma$ while $\theta$ is
fixed to the experimentally chosen value
in~\cite{Hensinger+01a}, \textit{i.e.}  $\theta=1.7$. When 
$\gamma$ is increased, the pairs of symmetric tori appear 
at~$\gamma\simeq0.56$. At the center of each set ot tori, there
is a periodic orbit. Over one period of the driving, the periodic
orbit is essentially a rotation over the fixed point at $(p,q)=(0,0),$
which explains that the whole structure remains trapped in a single
spatial cell.
For
$0.56\lesssim\gamma\leq\gamma_c,$ the tori remain nested in one
connected stable island. At~$\gamma=\gamma_c\simeq0.625,$ a chaotic
sea separates the symmetric islands which shrink and move away from
the central point before being dissolved through a cascade of
bifurcations starting at~$\gamma\simeq0.97$.
  
In one series of experiments, $\gamma\simeq0.96$ and~$\hbare\simeq0.8$, 
the atoms
are prepared in one island and their average momentum $\langle
p\rangle$ is measured stroboscopically every modulation period
($=2\pi$ in dimensionless units).  Since, in phase space, the islands
rotate about the origin with the same period, if no tunnelling occurred
no variation in $\langle p\rangle$ would be noticeable.  In fact,
starting the measurement sequence when $\langle p\rangle$ has its
maximum value, oscillations are observed which illustrate the back and
forth motion of the atoms between the islands due to dynamical
tunnelling.  The tunnelling period~$T$ is about 10 modulation periods
in this case ($200\ \mathrm{\mu s}$).  
This is in perfect agreement with the quasi-energy
splitting obtained numerically for
the two Floquet eigenstates having the largest Husimi functions
inside the islands.

It is worth noting that the
\textsc{nist} group uses a Bose-Einstein condensate as a preliminary
step for preparing atoms in well defined quantum states,
especially for achieving a large coherence length
for the wavefunction, i.e. a small spreading of the
Bloch angle $k.$ 
In order to prepare phase space localized states,
an optical lattice is carefully turned on. When the
tunnelling experiment starts,
the atomic density and the interaction between atoms is 
sufficiently small, and the experiment can be analyzed 
as the interaction of individual independent atoms 
with the laser beams, i.e. using Hamiltonian~\eqref{eq:ham}.
However, the cloud of atoms
remains cold enough, at a sub-recoil temperature, to prevent a 
large thermal
broadening of momentum distribution that would destroy the signal.
Because they start from very low temperature, these
preparation techniques based on condensate manipulation seem to allow
a wider room to manoeuvre than those working with thermal clouds only.
Adiabatic switching of the light potentials is not required and one
can actually work with values of the classical parameters~$\gamma$
and~$\theta$ which are far from the perturbative regime of an
integrable system.

\begin{figure}[!ht]
\center 
\includegraphics[width=8.5cm,keepaspectratio]{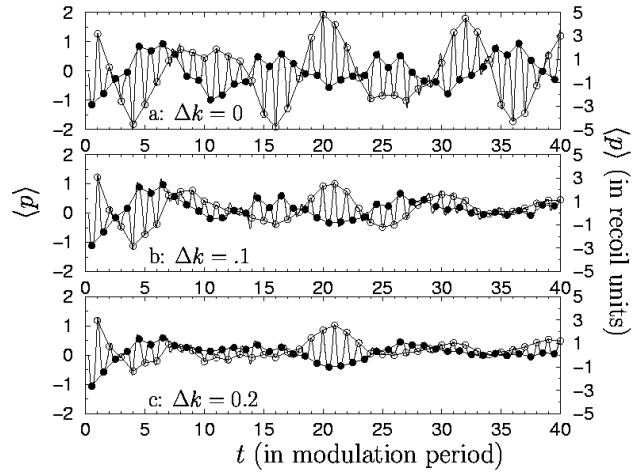}
\caption{\label{fig:pavbloch} Numerical simulation
of the quantum evolution, in the conditions of the
\textsc{Nist} experiment, i.e.
$\theta=1.72$ ,$\gamma=0.96$ and $\hbare=0.8$ (compare to Fig.~4~(a)
of~\protect\cite{Hensinger+01a}). 
Starting at~$t=0$ with a Gaussian
wave packet whose Husimi function is localized in one stable island
(with a vanishing average momentum), 
we follow the average momentum~$\langle p\rangle$ as time evolves. 
The stroboscopic measurements at
times~$\pi+2m\pi$ (resp. $3\pi+2m\pi$) with
$m\in\{0,1,\dots,40\}$, are plotted 
with the white (resp. black) circles. 
The tunnelling oscillations are clearly visible; the tunnelling period
can be extracted
from the typical time scale of the envelop: it is about 10 
modulation periods. In the upper plot (a),
we assume that a single Bloch angle $k=0$ is
initially prepared (which implies a perfect phase coherence of
the wavefunction
across the optical lattice).
The (thermal) dispersion of the Bloch angle washes out the signal: 
in case (b), we take a  momentum
distribution with width~$\Delta p=\alpha=2\Delta k=0.2$ and in case
(c) $\Delta p=0.4$. In the latter case, the amplitude of the envelopes
are so weak that this corresponds to an upper bound in temperature
(about 1/5th of the recoil temperature) 
at which tunnelling can be
measured.}
\end{figure}

By diagonalizing the evolution operator corresponding
to eq.~\eqref{eq:ham} over one period, we are not only able to reproduce
the oscillatory behavior of~$\langle p(t)\rangle$ 
(see Fig.~\ref{fig:pavbloch}~(a)),
but also we can study the spoiling effect of the thermal
dispersion~$\Delta p\propto\sqrt{\mathrm{temperature}}$ and predict
the maximum allowed temperature (see Fig.~\ref{fig:pavbloch}~(b) 
and (c)). \footnote{After the first version of the present paper was written, we  
learnt that an independent numerical work~\cite{Luter/Reichl02a} had obtained 
the same results as the ones in our Fig.~\ref{fig:pavbloch}~(a). But \cite{Luter/Reichl02a}
do not consider the thermal effects and work always within the Floquet theory at 
$k=0$ instead of the Floquet-Bloch theory. In the present work we clearly demonstrate
that thermal effects cannot be neglected and must be studied carefully  when experiments
are discussed.}  
If $\alpha$ denotes the width of the momentum
distribution in recoil momentum units, it can be
shown~\cite[\S6.a]{Mouchet+01a} that it corresponds to a statistical
mixture of Bloch states with $\Delta k=\alpha/2$. 
Fig.~\ref{fig:pavbloch}~(a)
corresponds to the ideal situation where all atoms are prepared
with $\alpha\ll 1$ about the $k=0$. 
When a small but non vanishing $\alpha$ is introduced, some states of
the quasi-energy bands with non vanishing~$k$ get involved and blur
the tunnelling oscillations. For
$\alpha=0.2$, the oscillation amplitude is reduced by a factor 2
and for~$\alpha=0.4$ have nearly disappeared. 
Therefore, in this experiment,
having a sub-recoil atom cloud is required.

In the following we want to focus on tunnelling only and  
we will implicitly keep $k=0$.

\subsection{Austin experiments~\protect\cite{Steck+01a}}
\label{subsec:Steck}

For a better understanding of the dynamics, one must go
beyond the two-level model involving the symmetric and the
antisymmetric states only. Other states must be taken into account and
their influence can be enlightened when a classical parameter
($\gamma$ or $\theta$) or the quantum one ($\hbare$) is continuously
varied.  Two (quasi-)energies may exactly get degenerate if they
belong to distinct symmetry classes. If not, they may follow a
so-called avoided crossing whose size 
reflects the direct coupling between
the two states (more precisely the off-diagonal matrix element of the
coupling perturbation), but also
the indirect coupling with other states. One of the keys of the
chaotic tunnelling problems is to identify clearly the qualitative
nature and the quantitative influence of indirect coupling.  This is
the background of Austin experiments.

\begin{figure}[!ht]
\center  
\includegraphics[width=8.5cm,keepaspectratio]{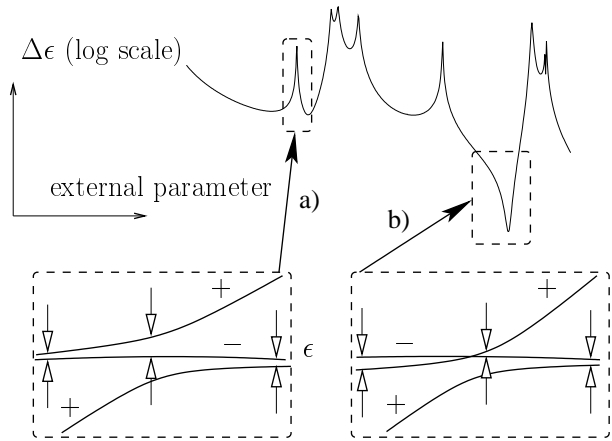}
\caption{\label{fig:fluctuation3levels} 
When a third level is crossing a tunnelling doublet when a
parameter of the Hamiltonian is varied, there is an avoided
crossing between the third level and the member of
the doublet with the same symmetry, while the other member of the
doublet (with opposite symmetry) ignores the third level.
Two generic scenarii exist:
in case (a), the tunnelling splitting increases in the vicinity of the
avoided crossing; in case (b), it decreases
and vanishes at a specific value of the parameter.   
}
\end{figure}

\begin{figure}[!ht]
\center  
\includegraphics[width=8.5cm,keepaspectratio]{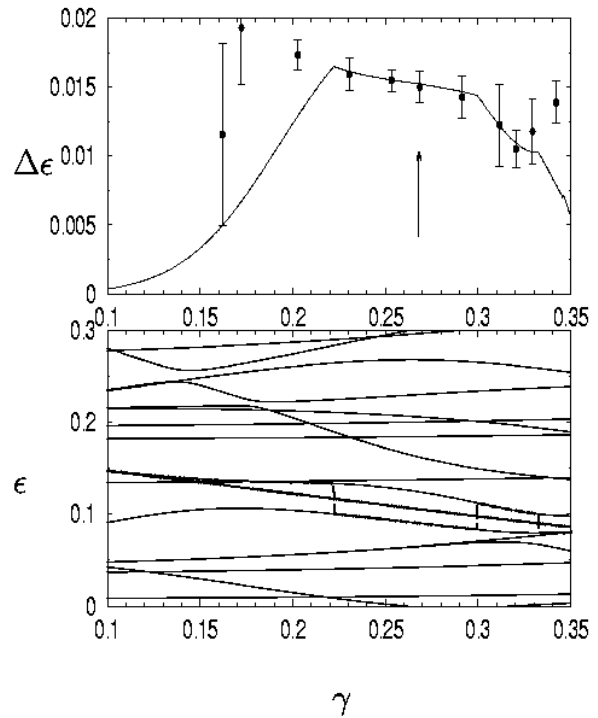}
\caption{\label{fig:austinqesplit}Numerical results obtained
with the parameters of the Austin experiment,
that is Hamiltonian~\eqref{eq:ham}, with $\theta=1$ and $\hbare=0.33.$
The lower plot shows a part of the quasi-energy spectrum when~$\gamma$ 
is varied.
Thick lines show the two quasi-energies whose difference is the tunnelling 
splitting plotted in the upper plot.
The two states are selected to have the largest localization of the
Husimi function 
at the center of the stable islands. When a third
level couples to the state that belongs to the same symmetry class,
 an avoided crossing can be seen and
the definition of the doublet becomes necessarily ambiguous. 
There, some discontinuity in the selected state 
(and in the slope of the tunnelling frequency) cannot be avoided.
The tunnelling splitting is shown as a function of $\gamma$ in the
upper plot, and is compared with the experimental results
from ref.~\protect\cite{Steck+01a}. The agreement is very good, which validates
the numerical approach. Around $\gamma=0.2$, a discrepancy is visible. This
is precisely the ``ambiguous" region where the dynamics cannot be reduced
to a simple tunnelling oscillation, 
but at least three levels must be taken into
account, leading to several relevant energy splittings.} 
\end{figure}

\begin{figure}[!ht]
\center  
\includegraphics[width=8.5cm,keepaspectratio]{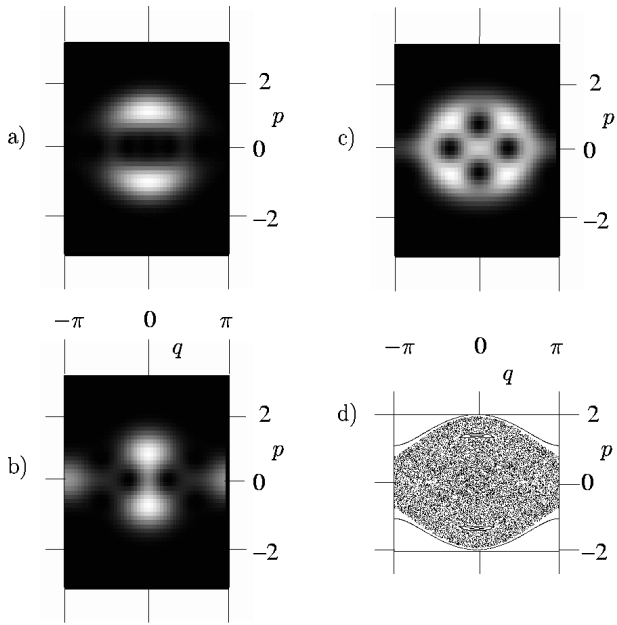}
\caption{\label{fig:austinhusimi}
 (a), (b) and (c) show the density in gray scale
of the  Husimi functions associated with the three states which
play an important role for tunnelling at~$\gamma=0.25$ (parameters
of the Austin experiment, as in fig.~\protect\ref{fig:austinqesplit}. 
As expected, the two members of the doublet, (a) and (b) have
their Husimi function localized about the two symmetric islands visible
in the classical Poincar\'e surface of section (d). (b) is strongly 
coupled to a third states shown in (c).
$\hbare$ is too large to attribute 
any regular/chaotic character to the third state. (c) clearly plays
a role in the enhancement of the tunnelling splitting through an
indirect coupling; this is thus a ``assisted tunnelling" mechanism,
which cannot be unambiguously characterized as chaos assisted.}
\end{figure}

A third level is involved in a non
negligible indirect coupling when its quasi-energy approaches the
tunnelling doublet energies. This can be understood from perturbation
theory as the leading term of the indirect coupling is
proportionnal to the inverse of the energy difference.  The two
generic scenarii of the crossing of the doublet by a
third state are shown in Fig.~\ref{fig:fluctuation3levels}.
Aside from the
unavoidable ambiguous definition of the splitting, it appears clearly
that case a) corresponds to an increase of the tunnelling splitting
during the
crossing while, conversely, case b) can lead to arbitrary small
splittings since an exact degeneracy occurs.  Thus, such a crossing by
a third state produces a sharp variation of the tunnelling period that can
be measured experimentally. This is actually what is observed in
the Austin experiments and can be confirmed by  numerical
experiments as shown in Fig.~\ref{fig:austinqesplit}.  
When looking at the Husimi
representation of the states, Fig.~\ref{fig:austinhusimi}, 
one can immediately distinguish between
the tunnelling doublet and the third state sufficiently far
from  the crossing.
As expected, the tunnelling doublet has Husimi representations
localized in the stable islands though they also spread in the
chaotic sea.  On a classical Poincar\'e surface of section, it is
easy to make the difference between chaotic and regular motion; how
to transpose this distinction at the quantum level is not known with the
large values of~$\hbare$ used in both the Austin and the
\textsc{Nist} experiments.
Some classical structures much smaller than the de Broglie
wavelength are possibly present in some of the states
\footnote{From time to time, it is claimed~\cite{Zurek01a} that
a simple matching between classical structures and quantum wave
sub-Planckian structures was found but, as it was understood long time ago
by~\cite{Balazs/Voros90a}, the latter are generally washed out 
as soon as one try to
measure global averages.}, but  just looking at the
Husimi representation of the third state for $\hbare\simeq1$
does not make it possible
to attribute any chaotic or regular character to
it. It is only for much smaller~$\hbare$ that chaotic
or regular wavefunctions make sense.
 This is not surprising: the
dichotomy between regular and irregular motion is a
classical one and, at present, it can be extrapolated at the quantum level
within the semiclassical regime only. Anyway, one must keep in mind that 
tunnelling does only make sense in the semiclassical regime as 
well.
  One of the great merit of
Austin experiments is to show for the first time a quantum
tunnelling effect where an indirect process is involved, but 
we fill exagerated to attribute any chaotic origin to it.

\section{Numerical evidence of a chaotic tunnelling regime}
\label{sec:numexp}

It is one of the first success of quantum chaos to have shown that the
energy levels of an integrable system ignore each others because
they are localized on different classical tori, while in
chaotic systems level repulsion is the rule. Following the discussion
in the previous section, one may therefore expect that the average
size of avoided crossings is increased when chaos is present.  
The fluctuations
of a tunnelling process which should be narrow and sparse in an
integrable regime should be broader, more numerous and possibly
involving many states in a chaotic regime. We now illustrate
this statement in the framework of the
experimental atomic Hamiltonian~\eqref{eq:ham}.

There are two different ways of rendering chaos observable by quantum
eyes: the first one consists in increasing the volume of the chaotic sea
with the help of a classical parameter, the other one
fixes the classical dynamics and
decreases~$\hbare$. We will present both ways.

\textbf{CAT, RAT and all that...}  
One may try to study the tunnelling
\emph{fluctuations} separately from the \emph{average
behavior}.  This average --- in a somewhat vague sense --- is
increased because chaos diminishes the classical dynamical
barrier\footnote{The question of defining an integrable system with
respect to which the chaotic one should be compared is extremely
difficult because of the exponential sensitivity of tunnelling on
classical parameters. It is a much more serious problem than the way
of defining an averaging procedure.} and this is the reason why the
phenomenon can be called chaos assisted tunnelling. However, as far
as only fluctuations are concerned (there can be an enhancement or a
decrease as well), the words ``chaos assisted tunnelling'' (CAT)~\cite{Tomsovic/Ullmo94a} may
lead to confusion and we simply use
``chaotic tunnelling''.  At last, Brodier, Schlagheck and Ullmo discovered what
they called  ``resonant assisted tunnelling'' (RAT)~\cite{Brodier+01a} to describe an
enhancement of tunnelling due to an indirect process which involves
one or several quasi-modes localized in the secondary resonances
surrounding the main symmetric islands.  Up to now, RAT 
has been studied quantitatively in a purely
quasi-integrable case but there are clues that it
can be extended far beyond \textsc{kam} theory.

\subsection{$\gamma$ change}

Let us first take $\hbare$ ten times smaller than in the \textsc{Nist}
 and Austin
experiments. It is much easier to do
it numerically than experimentally, as it requires to increase the
modulation frequency of the laser beams by one order of magnitude. 
We then
follow the quantum states through the classical bifurcation 
shown in picture~\ref{fig:bifurc2yeuxclassique} and discussed in
section~\ref{subsec:Hensinger}.  
Again, in order to calculate the
tunnelling splitting $\Delta\epsilon$, we select the states that have
the largest  Husimi functions inside the islands.

\begin{figure}[!ht]
\center  
\includegraphics[width=8.5cm,keepaspectratio]{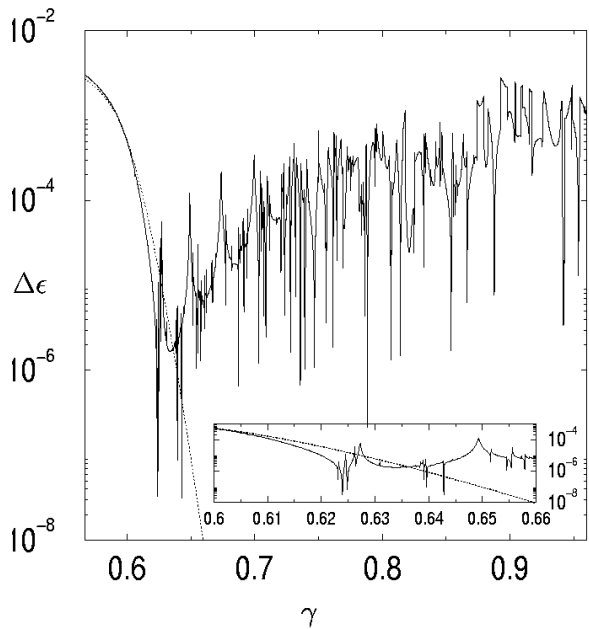}
\caption{\label{fig:split_flogamma}The tunnelling splitting as a function
of $\gamma$, for~$\theta=1.724137$ and~$\hbare=0.079638$, 
that is 10 times smaller than in the \textsc{nist} experiments.
Two regimes are clearly
 separated by the critical value~$\gamma_c\simeq0.625;$ these 
 two {\em quantum} regimes
 correspond to two different {\em classical} regimes
 in the Poincar\'e surfaces of section in
 Fig.~\ref{fig:bifurc2yeuxclassique}. 
  The smooth average decrease with sparse and narrow fluctuations  
  ($\gamma<\gamma_c$)
corresponds to the case where the symmetric classical
tori belong to the same  regular island. The Hamiltonian can there be approximated
by a simple integrable Hamiltonian, using the normal form
described in the appendix. The dotted line
 corresponds to the splitting calculated with this normal form
 and is in good agreement with the numerical result.
 In the second regime ($\gamma>\gamma_c$),
there are
 huge quantum fluctuations of the tunnelling splitting
 (and a slightly increased average value). Classically,
 this corresponds to tunnelling between unconnected symmetric islands,
 separated by a chaotic sea.}
\end{figure} 

\begin{figure}[!ht]
\center  
\includegraphics[width=8.5cm,keepaspectratio]{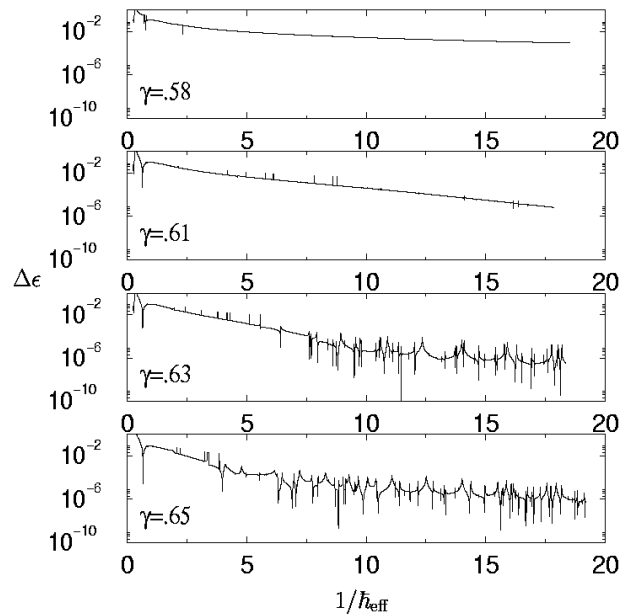}
\caption{\label{fig:split_flohbar}The tunnelling splitting as 
a function of $\hbare$, for fixed values of the
classical parameters, i.e. fixed classical dynamics
($\theta=1.724137$ and four values of $\gamma$). 
In the regular regime, an exponential decrease, as described by
eq.~\eqref{eq:wilkinson} is visible, as a straight line 
with a negative slope in a logarithmic scale.
The regime of fluctuations 
can be seen for~$\gamma>\gamma_c$ when $\hbare$ is small enough 
for the de Broglie wavelength
to be comparable to the size of the chaotic sea between the two islands.
}
\end{figure}
Figure~\ref{fig:split_flogamma} shows that, after a smooth
decrease of $\Delta\epsilon$ up to~$\gamma=\gamma_c\simeq0.625$
there is an abrupt change of regime. First, the mean value of
$\Delta\epsilon$ increases and second, many fluctuations appear 
which modify the
splitting by several orders of magnitude.  
It is remarkable 
that this change of regime can be matched on the Poincar\'e surfaces of
section. The smooth tunnelling regime occurs
the resonant tori belong to one quasi-integrable island. 
For~$\gamma<\gamma_c,$ we are able to reproduce the main
features of tunnelling by using an integrable approximation (see appendix).
The decrease of tunnelling is directly understood in terms of the 
lengthening of the
dynamical barrier.  $\gamma_c$ corresponds exactly to the point where
the islands get disconnected.  In addition, one can  follow the
bifurcation on the Husimi representations  of the
tunnelling doublet.

One can hardly detect by eye any regularity in the chaotic regime but
four large spikes in the range~$\gamma_c<\gamma<0.7$. This can be
traced back to the crossing by the same third state whose quasi-energy
line is folded four times in the Floquet zone centered on the doublet.

\subsection{$\hbare$ change} 

Figure~\ref{fig:split_flohbar} shows~$\Delta\epsilon$ as a function of~$1/\hbare$
for four values of~$\gamma$.  Here again, two types of regimes, a
quasi-regular and a chaotic one, can clearly be distinguished.  For
the values of $\gamma$ where some substantial chaos is present in
between the islands, by decreasing~$\hbare,$ one gets to the chaotic
regime.   For $\gamma<\gamma_c$, \textit{i.e.} in the
quasi-regular regime, one may enter in a chaotic regime but for far
much low value of~$\hbare$ since the chaotic layers between
\textsc{kam} tori are so thin that they are not even resolved in the
Poincar\'e surface of section given in the figures.
 
On this plots, a purely exponential law given by eq.~\eqref{eq:wilkinson}
would produce a straight line with negative slope. It gives rather
poor
predictions in the chaotic regime and should be corrected even in the
quasi-integrable regime in order to reproduce the
fluctuations. Whenever such a fluctuation is due to a crossing by a
quasi-mode (and not a state delocalized in the surrounding chaotic
sea), it might be reproduced by the resonance assisted tunnelling
techniques.

\section{Statistical signature of chaotic tunnelling ?}
\label{sec:subtleties}

In order to reproduce quantitatively the statistics of the tunnelling
splitting fluctuations in the chaotic regime, Leyvraz and Ullmo~\cite{Leyvraz/Ullmo96a}
have
introduced a random matrix model. The Hamiltonian
can be split in two uncoupled components associated
with the two even and odd symmetry subspaces. The corresponding
matrices are written as:
\begin{equation}
H^{\mathrm{even}}=
	\begin{pmatrix}\epsilon_0^+  & v^+_1       & v^+_2 & \cdots\\
                        v^+_1        &             &       &       \\
 		        v^+_2        &             &  H^+_\perp      &       \\
 		        \vdots       &             &       &       
	  \end{pmatrix}
\end{equation}
and
\begin{equation}
H^{\mathrm{odd}}=	 
	  \begin{pmatrix}\epsilon_0^-  & v^-_1       & v^-_2 & \cdots\\
                        v^-_1        &             &       &       \\
 		        v^-_2        &             &  H^-_\perp      &       \\
 		        \vdots       &             &       &       
	  \end{pmatrix}
\end{equation}
where $\epsilon_0^{\pm}$ represent the energies of the doublet,
$H_{\perp}^{\pm}$ the Hamiltonian in the chaotic sea (modelled by
a random Gaussian matrix) and $v$ the indirect coupling.

\begin{figure}[!ht]
\center  
\includegraphics[width=8.5cm,keepaspectratio]{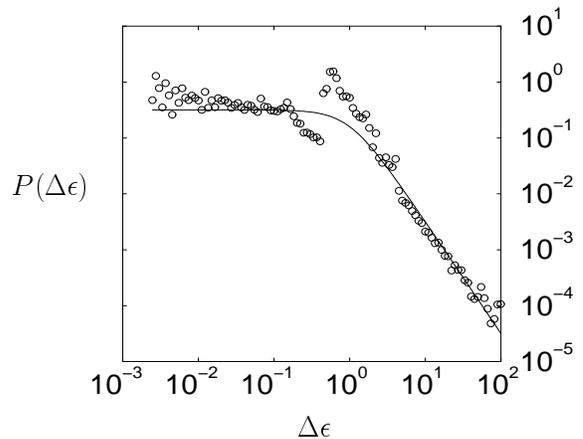}
\caption{\label{fig:LUlawcounterexample} Statistical distribution 
of the tunnelling splittings as $\hbare$ is varied in the quasi-integrable case
corresponding to Fig.~\ref{fig:2well_generalization1}~(b) ($\theta=1$, $\gamma=0.018$).
Surprisingly,
the numerically observed distribution (small circles)
is in good agreement with the Leyvraz-Ullmo law (solid line) which
is supposed to be valid in the chaotic case, as it 
treats all the states coupled to the tunnelling doublet on the same footing.
This clearly indicates that a Leyvraz-Ullmo distribution
is not sufficient to characterize chaotic tunnelling.
}
\end{figure}

Neglecting direct tunnelling consists in taking~$\epsilon_0^+=
\epsilon_0^-$.  The central hypothesis is to consider all the~$v$'s as
independent variables with \emph{the same} Gaussian distribution. This
is quite natural to treat all the other states on the same footing,
as they are assumed to be chaotic states randomly delocalized in
the chaotic sea.  With these assumptions, the splitting
distribution can be calculated 
and is given by a (truncated) Cauchy distribution, see~\cite{Leyvraz/Ullmo96a}.  
For
Hamiltonian~\eqref{eq:ham}, in each chaotic case where it has been
tested, the Leyvraz-Ullmo prediction is in agreement with the
numerical results (see~\cite{Mouchet+01a}).  More surprisingly,
we have found that the Leyvraz-Ullmo law gives correct predictions
even when the classical dynamics is quasi-integrable
(see Fig.~\ref{fig:LUlawcounterexample} and the corresponding Poincar\'e
surface of section in Fig.~\ref{fig:2well_generalization1}) and
the de Broglie wavelength  much larger than the chaotic
layers.
  When looking at the splitting, one observes numerous and
large fluctuations (over several orders of
magnitude) which were supposed to characterize the chaotic regime.
Conversely, we have not been able to find a regular regime of
fluctuations with a classically chaotic dynamics.  
Therefore one must conclude that even though two different
regimes of tunnelling fluctuations can be identified unambiguously,
classical chaos appears to be a sufficient but not a necessary
condition for having numerous and large fluctuations governed by the
Leyvraz-Ullmo law.

These unexpected results are not explained at the present state of the
theoretical approaches of chaotic tunnelling. If it appeared, within
future numerical or real experiments, that these results are not due
to the peculiar properties of our system, it would definitely mean
that theoretical studies should be all the more needed.

\section{Conclusion}
\label{sec:conclusion}  

In this paper we have studied in great details the transition between
a regular and a chaotic regime of tunnelling within a classical
configuration that can be achieved experimentally. We have shown why
$\hbare$ has to be small enough if one wants to reach for the first
time the chaotic regime in real systems. In  recent experiments
with cold atoms, it requires to increase the modulation period up to
one order of magnitude (that is at least the~MHz).

Theoretically, there is also a lot of work to do if we want to
understand and therefore predict the fluctuations quantitatively. The
semiclassical regime requires to study carefully the dynamics with
complex coordinates. However, the present work has shown clearly 
that the abrupt transition between a regular and a chaotic regime of tunnelling
corresponds to a classical transition that can be identified very precisely. 
Tunnelling being a relevant concept in a semiclassical regime only,
 it is therefore not surprising that the future investigations on chaotic 
tunnelling will have to 
keep track of the classical dynamics in one way or another.

\begin{acknowledgments} 
We acknowledge A. Shudo, S. Tomsovic, D. Ullmo
and W. Hensinger for stimulating discussions.  A. M. is grateful
towards O. Boebion for computer assistance in the Laboratoire de
math\'ematiques et de physique th\'eorique of Tours and thanks the
Laboratoire Kastler Brossel of Paris for kind hospitality.
Laboratoire Kastler Brossel de
l'Universit\'e Pierre
et Marie Curie et de l'Ecole Normale Sup\'erieure is
UMR 8552 du CNRS. Laboratoire de math\'ematiques et de physique th\'eorique
de l'Universit\'e Fran\c{c}ois Rabelais is
UMR 6083 du CNRS.
\end{acknowledgments}

\appendix
\section{Integral approximation near the pitchfork bifurcation}
When one classical parameter is smoothly varied in a non-integrable
Hamiltonian system, its periodic orbits follow infinite fractal-like
cascade of bifurcations through which they cannot be followed
smoothly. However, all the bifurcations can be classified according to a
simple set of scenarios: for a given bifurcation of a given periodic
orbit, the phase-space dynamics of the original parameter-dependent
Hamiltonian can be uniformly approximated by an integrable
parameter-dependent Hamiltonian which retains the relevant 
features only. Of course, because one cannot get rid of the chaotic
dynamics, this approximation does make sense only locally, that is
near the periodic orbit, in the neighborhood of the bifurcation. 
It is the object of the
Hamiltonian normal form theory to classify the bifurcations and obtain
the simplest form of the approximated integrable Hamiltonians (the
so-called normal forms).  Generically, \textit{i.e.} when no
constraint or symmetry is present, the one parameter Hamiltonian
normal forms have been completely classified by 
Meyer~\cite{Meyer70a,Meyer/Hall92a,Leboeuf/Mouchet99a}.
However, the
bifurcation shown in Fig.~\ref{fig:bifurc2yeuxclassique} is out of
the scope of Meyer's classification precisely because the time
reversal symmetry plays a key role. Even if the Poincar\'e surface of
section near the origin (see Fig.~\ref{fig:bifurc2yeuxclassique}~(b) cannot
be distinguished from the one corresponding to Meyer's transitional
bifurcation (see for instance Fig. 9~(b) in~\cite{Leboeuf/Mouchet99a}),
it is crucial to note that, in our case, two \emph{distinct}
$2\pi$-periodic orbits have emerged from the origin. In a transitional
bifurcation, the two stable islands would correspond to the
\emph{same}~$4\pi$-periodic orbit. Therefore they would be classically
connected to each other and would be irrelevant for tunnelling.  
Let us sketch briefly how to obtain the Hamiltonian
normal form in our case.

1. The first step is to find the value~$\gamma_0$ of~$\gamma$ at
which the bifurcation occurs ($\theta$ is kept fixed). Such a
bifurcation occur when the trace of the monodromy matrix~$M_\gamma$
at the origin after one period is 2 and corresponds in the $(-2\gamma,
4\gamma\theta)$-plane to the border of the even Arnold tongues (see
Fig. 20.1 in~\cite{Abramowitz/Segun65a}) (the transitional case occurs
when ${\mathrm{tr}}M_\gamma=-2$  at the border of the odd
Arnold tongues). More precisely, the case shown in
Fig.~\ref{fig:bifurc2yeuxclassique} corresponds to
$\theta=1.724137$ (the experimental value in the \textsc{Nist}
experiment) and the bifurcation takes place at
$\gamma_0\simeq0.564673$. In the following, we use
$\varepsilon=\gamma-\gamma_0$.

Let us denote by~${\mathcal{Y}}_\varepsilon(z)$ [resp.~${\mathcal{Z}}_\varepsilon(z)$]
 the solution of the Mathieu equation 
\begin{equation}
	y''(x)+(4\gamma\theta+4\gamma\cos(2x))y(x)=0
\end{equation}
such that~${\mathcal{Y}}_\varepsilon(0)=0$ and ${\mathcal{Y}}'_\varepsilon(0)=1$ 
[resp.~${\mathcal{Z}}_\varepsilon(0)=1$ and ${\mathcal{Z}}'_\varepsilon(0)=0$]. The prime
 stands for the derivative with respect to~$x$.  
For~$\varepsilon=0$, it is straightforward to show that
 \begin{equation}
	 M_{\gamma_0}=
	\begin{pmatrix}
	 1&\frac{1}{4\pi}{\mathcal{Z}}'_0(\pi)\\
	 0&1
	\end{pmatrix} 
\end{equation}
where ${\mathcal{Z}}'_0(\pi)\simeq1.480919$ for $\theta=1.724137$.
 
2. The second step is to make a (linear) $2\pi$-periodic canonical
change of coordinate that eliminates the time-dependence in the
quadratic part of Hamiltonian~\eqref{eq:ham}, near the origin and uniformly
in~$\varepsilon$. We are then led to the Hamiltonian
\begin{equation}
  \begin{split}
	\left(-\frac{1}{2}\frac{1}{4\pi}{\mathcal{Z}}'_0(\pi)+\alpha\varepsilon\right)q^2
	+\frac{1}{2}\beta\varepsilon p^2+\frac{1}{2}\delta\varepsilon p^2 
	\\ + \text{higher order $2\pi$-periodic terms}
 \end{split}      
\end{equation}

where~$\alpha,\beta,\delta$ are $\varepsilon$-independent
coefficients. Only
$\beta=\frac{1}{\pi}\partial_\varepsilon{\mathcal{Y}}_\varepsilon(\pi)$
evaluated at~$\varepsilon=0$ will be relevant since~$\alpha$ and $\delta$ 
can be
eliminated by a suitable canonical change of coordinates following a
method explained in~\cite[section 4.2]{Leboeuf/Mouchet99a}.  In our
case~$\beta\simeq2.008/\pi$.  We are therefore led to the following
normal form of the quadratic part of the Hamiltonian:
\begin{equation} 
  \begin{split}
	\left(-\frac{1}{2}\frac{1}{4\pi}{\mathcal{Z}}'_0(\pi)\right)q^2+\frac{1}{2}\beta\varepsilon p^2
	\\ 	+ \text{higher order $2\pi$-periodic terms} 
  \end{split}      	
\end{equation}
 
3. Following the same reasoning that lead to the transitional normal
form~\cite[sections and 4.2]{Leboeuf/Mouchet99a}, all higher order
$2\pi$-periodic terms, but the resonant terms of the
form~$h_k(\varepsilon)p^k$, can be cancelled by a suitable canonical change
of coordinates. Because of the time reversal symmetry, the
coefficient~$h_3$ vanishes identically and therefore the leading order
normal form is
\begin{equation}\label{eq:normalform}
	\left(-\frac{1}{2}\frac{1}{4\pi}{\mathcal{Z}}'_0(\pi)\right)q^2
	+\frac{1}{2}\beta\varepsilon p^2
	-\frac{1}{4} h_4(0) p^4.
\end{equation}
The explicit calculation of~$h_4(0)$ is tedious but it can be
estimated numerically by fitting the coordinate~$q=0,p=\pm
\sqrt{\beta\varepsilon/h_4}$ of the satellite $2\pi$-periodic orbits
for~$\varepsilon>0$. We obtain~$h_4\simeq0.0320$.

It is far from obvious that the quantization of the normal
form~\eqref{eq:normalform} will give a good approximation of the
quasi-energies of the tunnelling doublet. Some discrepancies may rise
from the fact that quantum physics is invariant under canonical
transformations only at the leading order in~$\hbar$.  When we
swap~$p$ with~$q$ and change the sign of the energies, the normal
form~\eqref{eq:normalform} leads to a standard 1D double-well problem
whose quantum spectrum can be found by numerically diagonalizing the
Hamiltonian written in a harmonic basis. The tunnelling splitting of
the ground doublet is given by the dotted line in
Fig.~\ref{fig:split_flogamma} and it agrees reasonnably well
with the exact numerical result
for~$\gamma$ near~$\gamma_0$.

\end{document}